\documentstyle[10pt,aaspp4]{article}
\begin{document}
\title{Radio Emission from Supernovae}
\author{J. I. Katz \\ katz@wuphys.wustl.edu}
\affil{Department of Physics and McDonnell Center for the Space Sciences \\
Washington University, St. Louis, Mo. 63130}
\authoremail{katz@wuphys.wustl.edu}
\begin{abstract}
I consider radio emission from the remarkable SN1998bw.  $^{56}$Ni and
$^{56}$Co decays produce a gamma-ray flux whose Compton-scattered electrons
naturally explain the observed mildly relativistic expansion of the radio
source and its double-peaked history.  Such models require a surrounding
plasma, perhaps produced by the supernova progenitor, whose interaction with
the nonrelativistic debris may account for the observed X-ray source.  The
radio spectrum appears to be self-absorbed.  This interpretation determines
the brightness temperature, and hence the energy of the radiating electrons,
implying a surprisingly large magnetic field.  Attempts to
avoid this conclusion by interpreting the spectrum as the result of inverse
bremsstrahlung absorption do not lead to significantly lower fields.  The
large inferred field may have several explanations: radiation from a central
pulsar, a turbulent hydrodynamic dynamo or an aspherical Compton current,
but a frozen-in field from the supernova progenitor is not adequate.  The
electron-ion and particle-field equipartition problems are discussed.
Compton electrons also explain the inferred expansion speed of SN1987A's
spots.
\end{abstract}
\section{Introduction} 
The unusual and peculiar SN1998bw was (\cite{K98}) the most luminous radio
supernova ever observed.  Its visible properties were also extraordinary
(\cite{G98}); a reported expansion speed approaching 60,000 km/s is larger
than that of other SN.   Modeling (\cite{I98}) indicates a
kinetic energy of $\sim 3 \times 10^{52}$ ergs, more than an order of
magnitude greater than that of most supernovae.  

Perhaps most remarkable was the apparent association of SN1998bw with the
unusual GRB980425, leading to the suggestion (\cite{B98}) of a new class of
supernova-gamma-ray burst.  The statistical significance and reality of this
association have been disputed (\cite{GLM99}), but the supernova is
extraordinary even without an association with a GRB.

\cite{K98} have argued on the basis of the absence of interstellar
scintillation and bounds on the radio brightness temperature that the radio
source of SN1998bw expanded semi-relativistically, with a Lorentz factor
$\Gamma$ in the range 1.6--2.  This value may be a clue to understanding
SN1998bw and its radio emission.  \cite{WL98} argued that the expansion was
in fact much slower, but this has not been generally accepted, and I will
adopt, at least approximately, the estimates of \cite{K98}.

A possible origin of the semi-relativistic expansion as a result of electrons
Compton-scattered by $^{56}$Ni and $^{56}$Co gamma-rays is suggested in \S 2.
In \S 3 I estimate the magnetic field in the source region implied by the
synchrotron self-absorption interpretation of the radio spectrum of
SN1998bw and discuss the origin of the field.  In \S 4 I mention possible
mechanisms for achieving electron-ion and particle-field equipartition.  In
\S 5 I ask if the radio spectrum may in fact be the result of inverse
bremsstrahlung by a thermal gas; although possible, this does not lead to
drastically lower estimates of the field than those presented in \S 3.  \S 6
questions the association of SN1998bw with GRB980425.  \S 7 contains a
summary discussion and predictions.
\section{Compton Electrons}
\cite{I98} inferred, on the basis of its visible light curve, that SN1998bw
contained about $0.7 M_\odot$ of $^{56}$Ni, or $N_{56} = 1.5 \times 10^{55}$
atoms.  The 6.10 d half life of $^{56}$Ni (an $e$-folding decay time of 8.8
d) is comparable to the duration of the first peak of radio intensity found
by \cite{K98}.  The daughter nucleus $^{56}$Co decays with a half life of 77
d (an $e$-folding decay time of 111 d), comparable to the rise and decay
time scale of the second, lower frequency, peak of radio emission in
SN1998bw.

The two characteristic time scales of radio emission and the double-peaked
intensity observed at some wavelengths suggest two distinct, but analogous,
processes.  In addition, the observation of expansion with Lorentz factor
$\Gamma$ in the approximate range 1.6--2 calls for explanation.  Is there is
a natural explanation of the observation of mildly relativistic expansion
with this particular value of $\Gamma$?

The decay of $^{56}$Ni (which is by electron capture) produces gamma-rays of
0.788 MeV, 0.812 MeV and 1.56 MeV, which are emitted in 48\%, 85\% and 14\%
of the decays, respectively (\cite{LHP67}).  The decay of $^{56}$Co, 80\% by
electron capture, produces gamma-rays of several energies, of which the
most important are 0.847 MeV (100\%), 1.04 MeV (15\%), 1.24 MeV (66\%), 1.76
MeV (15\%), 2.02 MeV (11\%), 2.60 MeV (17\%) and 3.26 MeV (13\%); 20\% of
its decays are by e$^+$ emission, with an energy distribution extending up
to 1.5 MeV.  This copious production of gamma-rays offers a possible
explanation of the existence and expansion speed of the radio source and
perhaps of its magnetic field.

It is necessary to assume that the $^{56}$Ni is mixed to the surface of the
debris; perhaps a jet or plume of $^{56}$Ni-rich material penetrates any
envelope and is expelled.  The absence of H or He in the spectrum of SN1998bw
establishes the absence at least of a massive envelope cloaking the high-Z
material.   After $10^6$ s at the observed photospheric
velocity of $6 \times 10^9$ cm/s this matter is in a shell of column density
$\approx 3$ gm/cm$^2$.  If mixed with an equal quantity of other material, to
make up a Chandrasekhar mass of debris, the total column density is $\approx
6$ gm/cm$^2$, which is transparent (optical depth $\approx 0.5$) to 0.812
MeV gamma-rays, the most important emission of $^{56}$Ni.

After the gamma-rays emerge from the supernova debris they enter any
surrounding medium, moving outward at the speed of light.  Their source
gradually becomes transparent to them as it expands, and their emergent
intensity increases with a rise time of a few days, until the competition
between increasing transparency and radioactive decay leads to a peak at
$\approx 10$ d, corresponding to the first peak in radio intensity.
Averaging over the Klein-Nishina cross-section, the mean electron produced
by Compton scattering of a 0.812 MeV gamma-ray has $\gamma = 1.65$ and the
most energetic has $\gamma = 2.12$.  The stopping column density of a
$\gamma = 1.65$ (kinetic energy $E_C = 330$ KeV) electron in dilute ($n_e
\sim 10^4$ cm$^{-3}$; the dependence on $n_e$ is only through the Coulomb
logarithm and is very weak) ionized plasma is $\ell_s \approx 8 \times
10^{21}$ cm$^{-2}$ (\cite{L80}).  Comparing this to the Klein-Nishina
cross-section $\sigma_{KN} = 2.6 \times 10^{-25}$ cm$^{-2}$ yields an
estimated efficiency of conversion of gamma-rays to Compton electrons
$\epsilon = \ell_s \sigma_{KN} \approx 2 \times 10^{-3}$; the remaining
energy appears as thermal heating of the plasma by Compton electrons which
stop within it.

If there is less than a stopping length of low-Z medium surrounding the
$^{56}$Ni then the efficiency will be reduced because in traversing the
high-Z debris the Compton electrons suffer Coulomb scattering by the
nuclei as well as energy losses to the electrons; the scattering length in
pure $^{56}$Ni is only $\approx 10^{19}$ cm$^2$ (\cite{S62}).  A
hydrogen-rich shell of column density $10^{22}$ cm$^{-2}$ (requiring only $4
\times 10^{-3} M_\odot$ at a radius of $6 \times 10^{15}$ cm, the distance
traveled by the massive debris in $10^6$ s) would be sufficient to
regenerate the full flux of Compton electrons and to restore the efficiency
to the value of the preceding paragraph.  As discussed above, supernovae are
expected to be surrounded by the remains of the winds of their progenitors.

These Compton electrons move outward with a mean speed $v_e \approx 0.8 c$
and current density ${\vec j}_{Compt}$.  Once they have expanded to a radius
$> 2$ times that at which they were born they move essentially radially
outward, which may be shown either by considering them as an adiabatically
expanding gas (non-radial components of motion are, in effect, random
thermal velocities, soon converted to ordered outward motion) or as
individual ballistic particles.  This free expansion requires a denser
ambient medium in order to provide a return (counter-) current density
${\vec j}_{cc}$ to maintain electrical neutrality.  The free-streaming
density of Compton electrons
$$n_{Ce} = {\epsilon N_{56} \nu_{56} \exp{(-\nu_{56} \Delta t)} \over 4 \pi
r^2 v_e}, \eqno(1)$$
where $\nu_{56}$ is the radioactive decay rate ($1.3 \times 10^{-6}$
s$^{-1}$ for $^{56}$Ni) and $\Delta t$ the time between creation of the
$^{56}$Ni and the decays whose Compton-scattered electrons are observed; the
nonrelativistic motion of the $^{56}$Ni and the gamma rays' path between
production and Compton scattering are neglected.  Adopting $\Delta t = 7$ d
and $r = 3 \times 10^{16}$ cm yields $n_{Ce} \approx 3 \times 10^4$
cm$^{-3}$, corresponding to $\approx 3 \times 10^{-3} M_\odot$ of hydrogen,
similar to values estimated elsewhere in this paper.  A medium even slightly
denser than this, such as plausibly produced by loss of the progenitor's
envelope over $\sim 10^3$ y, would supply the countercurrent required by
charge neutrality with only a small cost in energy to the Compton electrons
(the potential required to drive the countercurrent retards the Compton
electrons, but only slightly if the countercurrent density is significantly
larger and velocity significantly less).

After the $^{56}$Ni decays, $^{56}$Co produces its own gamma-rays, Compton
electrons and positrons.  The combination of positrons and Compton electrons
permits a neutral mildly relativistic wind even in the absence of a
background plasma to provide a countercurrent, should there be none.  In
other respects, the effects of $^{56}$Co decays are similar to those of
$^{56}$Ni decays, although the $^{56}$Co gamma-rays and Compton electrons
are more energetic (the abundant 1.24 MeV gamma-ray produces a mean Compton
$\gamma_e = 2.14$ and the gamma-ray spectrum extends up to 3.26 MeV).  This
new wave of Compton electrons will eventually outrun the $^{56}$Ni Compton
electrons and produce a second peak of radio emission, as observed.
\section{Magnetic Field}
\subsection{Self-Absorbed Spectrum}
Inspection of the radio spectrum of SN1998bw (\cite{K98}) shows evidence of
self-absorption.  The characteristic self-absorption frequency is $\approx 5$
GHz 10 d after the supernova, declines to $\approx 2$ GHz after 30 d, and
continues to decline as the source fades thereafter.  This accords with
expectations for an expanding self-absorbed synchrotron source.  The low
frequency (self-absorbed) flux first rises as the radiating area expands.
The higher frequency flux declines as a consequence of declining magnetic
field or electron energies.  It is possible to fit simple models to the data,
but the observed (\cite{K98}) double-peaked time dependence of the flux at
most frequencies implies that simple models will not be satisfactory.

The brightness temperature $T_B^\prime$ in the emitting frame may be
obtained directly from the observed intensity, if the size of the radiating
region is known, and \cite{K98} used such brightness temperatures, combined
with energetic arguments, to infer the rate of expansion of the
self-absorbed source:
$$k T_B^\prime(\nu^\prime) = {S_\nu d^2 \over 2 \pi b^2 t^2 \nu^2 D},
\eqno(2)$$
where $k$ is Boltzmann's constant, $S_\nu$ is the observed flux density, $d
= 38$ Mpc is the distance (assuming a Hubble constant of 65 km/s/Mpc), $\nu$
is the frequency of observation, $t$ is the elapsed time since expansion
began, $bc$ is the apparent velocity of expansion ($b = \Gamma \beta$ where
$\Gamma$ and $\beta c$ are the Lorentz factor and expansion speed), $D =
[\Gamma(1 - \beta \cos\theta)]^{-1} \sim \Gamma$ and $\nu^\prime = \nu/D$
where $\nu$ is the observed frequency and $\nu^\prime$ the frequency in the
comoving frame.

The brightness temperature of a self-absorbed source will generally
approximate the energy of the radiating particles (in thermal equilibrium it
equals the particle temperature).  Hence the radiating electrons
have a Lorentz factor in the co-moving frame
$$\gamma_e^\prime \approx {S_\nu d^2 \over 2 \pi b^2 m_e c^2 t^2 \nu^2 D}.
\eqno(3)$$
\subsection{Required Field}
The co-moving magnetic field $B^\prime$ may be estimated from (3) using the
synchrotron radiation relation $B^\prime \approx 2 \pi m_e c \nu^\prime/e
\gamma_e^{\prime\,2}$:
$$B^\prime = {8 \pi^3 m_e^3 c^5 \nu^5 b^4 t^4 D^2 \over e S_\nu^2 d^4}.
\eqno(4)$$

The values of $B$ inferred for SN1998bw are remarkably high.  For mildly
relativistic motion approximate $D \approx 1$, $b \approx \beta$ and $B
\approx B^\prime$.  The numerical values (\cite{K98}) $\nu = 2.49$ GHz
(nominally $\lambda = 13$ cm), $t = 1.01 \times 10^6$ s (11.7 d), $S_\nu =
19.7$ mJy and $d = 38$ Mpc yield
$$B \approx 0.13 \beta^4\ {\rm gauss}. \eqno(5)$$
These large inferred fields are a consequence of the low $T_b^\prime$ (and
$T_b$) implied by a rapidly expanding source of large linear size.  The low
brightness temperature combined with the interpretation of self-absorption
implies radiation by electrons of comparatively low energy, and hence a high
magnetic field is required to produce radiation of the observed frequency.

\cite{WL98} suggest $\beta \approx 0.3$.  This would imply a substantially
lower value of $B$ than for mildly relativistic motion (their numerical
estimates are somewhat larger than that of (5) because of differences in
various details).  \cite{K98} dispute such low values of $\beta$, and this
paper does not attempt to resolve this issue. 

Is the magnetic field (5) plausible, extended over a region of size $r
\approx \Gamma \beta c t \approx 3 \times 10^{16} b$ cm$^2$?  The implied
magnetic energy is
$${\cal E}_B = {D^2 B^{\prime\,2} r^3 \over 6} \sim 8 \times 10^{46}
\beta^{11} \Gamma^{11} D^4\ {\rm ergs}. \eqno(6)$$
This is modest if the expansion is not relativistic ($\Gamma \approx D
\approx 1$).  In that case \cite{WL98} point out that ${\cal E}_B$ is
several orders of magnitude less than the kinetic energy of SN1998bw and the
energy of the radiating electrons.  Only if $\Gamma$ exceeds 2 does ${\cal
E}_B$ approach the kinetic energy of the supernova debris.
\subsection{Sources of Field}
\subsubsection{Flux}
Energy is not the only relevant criterion.  If a magnetic field's
dependence on distance can be described by a power law $B \propto
r^{-\alpha}$ three simple models may be considered.  A static dipole field
has $\alpha = 3$; this is clearly inadequate.  If the field is frozen into a
conducting outflow either from the progenitor or from the SN itself, then
flux is conserved and $\alpha = 2$.  In this case the apparent inferred flux
$$\Phi_m \approx 10^{32} \beta^6\ {\rm gauss\,cm}^2; \eqno(7)$$
this flux is only apparent because it is $\int \vert {\vec B} \vert\,dS$
rather than $\int {\vec B} \cdot {\vec {dS}}$.  For comparison, the flux of a
typical pulsar is $\sim 10^{24}$ gauss-cm$^2$, that of a
(perhaps-hypothetical) ``magnetar'' is $\sim 10^{27}$ gauss-cm$^2$ and that
of the Sun is $\sim 10^{23}$ gauss-cm$^2$.  The magnetic fields of SN
progenitors are not directly measured but it is clear that the estimate (7)
is excessive, and that such a flux cannot be produced by a flux-conserving
flow.  If this model is to be salvaged there must be another source of field.
\subsubsection{Hydrodynamic Dynamo Fields}
When supernova debris runs into a surrounding medium the contact
discontinuity between the two shocked fluids is generally hydrodynamically 
unstable and may amplify pre-existing fields by a turbulent dynamo mechanism,
but this is not readily quantified.  Such a medium and shocks are not
necessary parts of a supernova model, but will occur if the supernova is
surrounded by the remains of a wind produced by its progenitor (\cite{B99}),
as is assumed elsewhere in this paper.  This is the only mechanism in which
the field is powered by the hydrodynamic energy of the supernova, although
only a small fraction of this energy is available unless the surrounding
medium is as massive as the fast debris.
\subsubsection{Radiation Fields}
Another possibility is the field of a propagating electromagnetic wave, for
which $\alpha = 1$.  Because the field alternates in direction with a short
wavelength the actual flux is not large, even though the field and apparent
flux may be large.  The magnetic field at a distance $r \gg c/\omega$ from a
source of rotational frequency $\omega$ and dipole moment $\mu$ is
$$B \sim {\mu \omega^2 \over c^2 r}. \eqno(8)$$
For a magnetic neutron star ($\mu \sim 10^{30}$ gauss-cm$^3$) rotating at
breakup ($\omega \sim 1.5 \times 10^4$ s$^{-1}$) $B \sim 10$ gauss at $r = 3
\times 10^{16}$ cm.  This is certainly more than sufficient, and allows for
smaller $\omega$ or $\mu$.  An electromagnetic wind from a neutron star (or
magnetized accretion disc) would be required to emerge through the dense
debris implied by the visible light curve of the supernova.  This is in
contrast to a GRB in which the presence of relativistic outflow implies the
absence of dense debris, at least in directions in which gamma-rays are
observed.  If the supernova debris is confined to dense filaments then the
electromagnetic wind may penetrate between these filaments, as in the much
older Crab Nebula.

If the magnetic field has its origin in the radiation of a new pulsar, then
(8) determines $\mu \omega^2$.  The spindown power is
$$P = {2 \over 3}{(\mu \omega^2)^2 \over c^3} \sim {2 \over 3}{B^2 r^2 c} \sim
10^{40}\ {\rm erg/s}. \eqno(9)$$
In principle, such a pulsar may be observed once the nonrelativistic debris
becomes transparent.

Eq.~(8) may be applied to other astronomical objects.  For example, a
similar estimate has been used to explain the field of the Crab nebula, in
which there is a pulsar of known properties, and no dense gas intervening
between it and the synchrotron emitting nebula.  Accretion discs around
black holes in AGN and extragalactic double radio sources are another
application; the rotating magnetized disc implies an oscillating magnetic 
dipole moment and radiation.  If the radiation is beamed into an
angle $\Omega$ then (8) should be replaced by
\begin{eqnarray}
B \sim {\mu \omega^2 \over c^2 r} \left({4 \pi \over \Omega}
\right)^{1/2} &\sim& \left({B_d \over 10^4\,{\rm gauss}}\right) \left({M
\over 10^8 M_\odot}\right) \left({r \over 300\,{\rm Kpc}}\right)^{-1}
\left({\Omega \over 10^{-2}\,{\rm sterad}}\right)^{-1/2}\ 3\,\mu{\rm gauss}\\
&\sim& \left({L \over 10^{46}\,{\rm erg/s}}\right)^{1/2} \left({r \over 300\,
{\rm Kpc}}\right)^{-1} \left({\Omega \over 10^{-2}\,{\rm sterad}}
\right)^{-1/2}\ 3\,\mu{\rm gauss},
\end{eqnarray}
where $B_d$ is the disc field and the last relation is obtained assuming
only that the disc viscosity is magnetic (\cite{K91}) and $L$ is the
accretional power.  In double radio sources equipartition is also suggested
if the pressure of the relativistic wind is balanced by that of the
electrons accelerated as its fields reconnect.
\subsubsection{Compton Current Fields}
Electrostatic neutrality requires ${\vec \nabla}\cdot{\vec j}_{Compt} = -
{\vec \nabla}\cdot{\vec j}_{cc}$ to high accuracy.  It does not require
${\vec j}_{Compt} = - {\vec j}_{cc}$, and if the flow is not spherically
symmetric the latter equality is not likely to hold.  As a result, the
Compton electrons may create closed loops of net current, with resulting
magnetic fields.  One characteristic value of the field is obtained from
Amp\`ere's Law: $B \sim \epsilon N_{56} \nu_{56} \exp{(-\nu_{56} \Delta t)}
e/(rc) \sim 10^{10}$ gauss!  It is evident on energetic grounds
that fields of this magnitude cannot be created; magnetic forces will induce
countercurrents which will cancel the divergence-free part of the Compton
current (as well as its divergence) to high accuracy.  However,
this cancellation will not be exact.  Just as electrostatically-driven
countercurrents leave a net potential (which drives them) of some fraction of
the Compton electron energy, magnetic countercurrents may leave a net current
sufficient to produce a field whose energy approaches equipartition with that
of the Compton electrons.  This energy is $\sim \epsilon N_{56}
\exp{(-\nu_{56} \Delta t)} E_c \sim 10^{46}$ ergs, corresponding to $B \sim
0.04$ gauss for $r \sim 3 \times 10^{16}$ cm.  This is comparable to that
suggested by the synchrotron self-absorption argument, and offers a possible
explanation of the apparent deviation of the magnetic energy from energetic
equipartition---equipartition is, in fact, achieved, but with the Compton
electrons' energy rather than with the bulk hydrodynamic energy.
\subsubsection{Plasma Instability Fields}
A related hypothesis attributes the magnetic field to electromagnetic plasma
instabilities resulting form the interpenetration of Compton- and
counter-currents.  In this case the field is chaotic on fine scales rather
than ordered, but the possible field energy is again comparable to (but
somewhat less than) the energy of the Compton electrons.
\section{Equipartition?}
\subsection{Particle-field Equipartition}
\cite{K98} assume equipartition between magnetic and particle energies in
their theoretical argument for mildly relativistic expansion.  There are two
classical arguments for this assumption.  The first is based on attempts to
calculate the generation of magnetic fields by dynamo mechanisms.  This
argument comes in as many forms as there are theories of dynamos, but it
generally concludes (or assumes) that when the magnetic energy density
becomes a significant fraction of the kinetic energy density its
back-reaction on the motion will suppress further dynamo activity.  If the
particle energy density is similarly limited by the hydrodynamic energy
density (\cite{K91}) then rough particle-magnetic equipartition will be
achieved.  Clearly, without detailed understanding of the dynamo and
acceleration processes in any particular configuration this argument is very
approximate, or perhaps only suggestive, but it does appear to be crudely
correct for the interstellar medium and (excluding the energetic particles)
for the Solar convective motion.

The second classical argument for particle-field equipartition notes that
for a given total energy the synchrotron power radiated will be maximized if
equipartition obtains.  Astronomical surveys are generally flux-limited,
implying that most detected sources will be fairly close to equipartition,
if it can be achieved.

It is unclear if either of these arguments applies to SN1998bw.  There may
be opportunity for Rayleigh-Taylor instability at the contact discontinuity
between the forward and reverse shocks when the debris encounters the
surrounding medium (likely a wind ejected by the SN progenitor; {\it cf.}
{\cite{B99}), but this depends on the details of the hydrodynamics.
Departures from spherical symmetry may lead to Kelvin-Helmholtz instability.
It is unknown if, or how effectively, these processes may produce dynamo
field amplification.  The radio counterpart of SN1998bw was detected by a
search targeted on this unusual visible and gamma-ray object, and was radio
flux-limited only in the implicit sense that its detection was limited by the
radio sensitivity.
\subsection{Electron-ion Equipartition}
Electron-ion equipartition presents a question entirely distinct from that
of particle-magnetic field equipartition.  All GRB models require at least
an approximation to electron-ion equipartition in order to couple the ion
kinetic energy to the electrons which radiate.  \cite{WL98} require this
too, and in fact their assumed velocity fairly approximates that required
(assuming a composition of helium or heavier elements) to produce their
estimated electron energies if equipartition occurs.  The mechanism of
electron-ion equipartition in a collisionless shock, relativistic or
non-relativistic may be as simple as an electrostatic double-layer with a
potential sufficient to slow the ions as implied by the shock jump
condition (\cite{K94}).

If cold electron and ion streams penetrate an orthogonal magnetic field
their differing gyroradii would lead to a charge separation which can only
be avoided by the presence of a potential sufficient to equalize the
gyroradii, which (in the relativistic limit) equally divides the kinetic
energy between electrons and ions (in the nonrelativistic limit the electron
kinetic energy exceeds that of the ions by the ion/electron mass ratio).
This is an oversimplification because it assumes the magnetic stress greatly
exceeds the hydrodynamic stress, but the qualitative justification for
equating gyroradii (and energies, in the relativistic regime)---that only if
the gyroradii are equal will electrostatic neutrality be maintained when 
shocks occur in magnetic fields---may be valid.  If the gyroradii are
unequal an electrostatic potential will develop which will equalize them.
\section{Inverse Bremsstrahlung?}
The interpretation of the low-frequency turnover in the radio spectrum of
SN1998bw as synchrotron self-absorption led to the inference of a magnetic
field too large to be easily explained.  An alternative explanation for this
turnover is absorption by inverse bremsstrahlung, as suggested by
\cite{C82}.  Inverse bremsstrahlung absorption will not, in general, produce
the quantitative $F_\nu \propto \nu^2$ or $F_\nu \propto \nu^{5/2}$ spectra
of synchrotron self-absorption, but when the low frequency turnover is only
defined by two spectral points of moderate accuracy, as was the case for
SN1998bw (\cite{K98}) it is impossible to distinguish between these two
explanations on spectral grounds alone.

If the low frequency turnover is the result of inverse bremsstrahlung
absorption then the observed brightness temperature is only a lower bound on
the brightness temperature in the (optically thin) emitting region.  This,
in turn, is only a lower bound on the energy of the emitting electrons.
Hence the magnetic field is only bounded from above, and may be as small as
required by a theory of field generation.  The actual field value depends on
the electron energy (or {\it vice versa}); in the absence of a detailed
theory of particle acceleration it is generally impossible to reject an
inferred value of the electron energy.  High electron energies need not,
however, imply excessive brightness temperatures because the source is
optically thin.  Hence, as argued by \cite{K98}, the synchrotron
self-Compton catastrophe may be avoided if the source is expanding
relativistically, which reduces the inferred brightness temperature, as
first pointed out by \cite{W66}.

If the magnetic field is too small then the ratio of inverse Compton
luminosity $L_{IC}$ to synchrotron $L_S$ becomes excessive.  The value of
$L_{IC}$ permitted by the observations depends on its (unknown) frequency,
but it probably safe to require it to be less than the visible luminosity
$L_V \sim 10^{43}$ erg/s, roughly $3 \times 10^4$ times the radio power at
$10^6$ s (\cite{K98}).  This sets a lower bound on $B$:
$$B \approx \left({2 L_V \over r^2 c}{L_S \over L_{IC}}\right)^{1/2} >
0.04\ {\rm gauss}, \eqno(11)$$
where the numerical value has assumed only mildly relativistic expansion.
The numerical result is close enough to the estimates derived for the
self-absorbed synchrotron model that discarding this assumption has not
materially reduced the difficulty of explaining the required field.

If the synchrotron source is expanding relativistically then $r \approx c t
\Gamma^2$ and the visible radiation intensity in the frame of the
synchrotron source is $\sim L_V/(4 \pi r^2 \Gamma^2) \propto \Gamma^{-6}$,
where the two extra powers of $\Gamma$ come from the redshift of the visible
radiation, and the lower bound on the comoving $B^\prime$ is reduced $\propto
\Gamma^{-3}$.  The bound on the laboratory frame $B$ is only reduced
$\propto \Gamma^{-2}$.  For mildly relativistic expansion, such as inferred
for SN1998bw, the required $B$ remains large.  In addition, because of the
larger inferred $r$, the various scaling laws yield estimates of of $B$
which are reduced by factors $\propto \Gamma^{-2\alpha}$.  The difficulty
is not resolved, and, in fact, is worsened for $\alpha > 1$.

The inverse bremsstrahlung opacity of hydrogen, including the effects of
stimulated emission, is
$$\kappa_{ff} = 3.69 \times 10^8 {n^2 h \over k T^{3/2} \nu^2} g_{ff}\ {\rm
cm}^{-1} \eqno(12)$$
(\cite{S62}).  Taking $T = 10^{4\,\circ}$K and $\nu = 2.49 \times
10^9$ Hz (for which $g_{ff} \approx 5$) yields $\kappa_{ff} \approx 1.2
\times 10^{-26} n^2$ cm$^{-1}$.  Effective absorption will be obtained in $3
\times 10^{16}$ cm if $n = 10^5$ cm$^{-3}$, requiring $\sim 10^{-2}$
M$_\odot$ of gas.  Such a circum-SN evelope, produced by a wind from the
progenitor star, is possible; if expelled at 10 km/s (as
appropriate to a red supergiant) its lifetime is $\sim 10^3$ yr.  Much more
massive winds are plausible.  Such winds have been widely discussed since
the discovery of ring structures around SN1987A, and have been inferred in
other supernovae ({\it cf.} \cite{B99}).

It is, of course, necessary that the envelope be ionized in order for inverse
bremsstrahlung to occur.  The envelope consists of $\sim 10^{55}$ atoms,
which require $\sim 2 \times 10^{44}$ ergs of ionizing ultraviolet radiation
to ionize.  The flux of SN1998bw in the Lyman continuum is not directly
observable, but the required energy is only $\sim 10^{-5}$ of the visible
radiation emitted, which is certainly plausible and will be found for a
Planck spectrum with $T > 9000^\circ$ K.  The resulting temperature of the
photoionized gas depends on the shape of the ultraviolet spectrum, but for a
comparatively cool spectrum it will be $\approx 10^{4\,\circ}$K, as assumed.

A dense and comparatively cool ionized gas is a source of recombination line
radiation.  A straightforward estimate using the preceding parameters and 
standard theory (\cite{O74}) leads to unobservably small ($\sim 10^{-5}$\AA)
equivalent widths, even for the Balmer lines, so that it is not possible to
test the hypothesis of inverse bremsstrahlung absorption in this manner.

As the ionized envelope is dispersed by the supernova debris (once
shock-heated its absorption coefficient decreases greatly because of the 
temperature dependence of $\kappa_{ff}$) the inverse
bremsstrahlung absorption will decrease.  This can qualitatively account for
the evolution of the radio spectrum, whose early deficiency of longer
wavelength flux gradually disappears.  Quantitative modeling might, in
principle, distinguish these predictions from those of self-absorption
models, but all models contain several free parameters; together with
the complexity of the observed radio spectral behavior (\cite{K98}) this
makes an unambiguous discrimination between the models difficult.
\section{GRB980425?}
GRB980425, if produced by SN1998bw, poses another problem.  The early states
of a stellar explosion would be expected to be very optically thick, and to
produce roughly a black body spectrum.  The observed spectrum (\cite{B98}),
if fitted to a black body, suggests a temperature $\sim 30$ KeV.  Combining
this with the inferred power (\cite{G98}) of $5.5 \times 10^{46}$ erg/s
implies an emitting radius (assumed spherical) of $7 \times 10^7$ cm.
However, the observed expansion speed of $6 \times 10^9$ cm/s (\cite{K98})
means that this radius will be exceeded about 0.01 s after the explosion
begins, inconsistent the observed GRB duration of about 10 s.  The power and
duration are consistent if the temperature is $\sim 1$ KeV, but this is
completely inconsistent with the observed spectrum.  

Matter expelled from a supernova core with temperature $\sim 30$ MeV at $r
\sim 6 \times 10^6$ cm will cool adiabatically to $\sim 3$ KeV, much less
than the value estimated from the observed spectrum, by the time it reaches
$6 \times 10^{10}$ cm, the radius at the period of gamma-ray emission.
In addition, a radiating photosphere will be much cooler than the temperature
deep within the outflow, for which adiabatic expansion is the only cooling
process.  The difficulty of explaining the observed properties of GRB980425
in the context of a supernova may best be resolved, despite the apparent
coincidence, if they are in fact unrelated events, as suggested on
statistical grounds (\cite{GLM99}).  This is supported by the
observation (\cite{P99}) of two transient X-ray sources, one long-lived and
coincident with SN1998bw and the other, briefer and not coincident with
SN1998bw, resembling a GRB X-ray afterglow.
\section{Discussion}
The most difficult part of the problem of the radio emission of SN1998bw is
explaining how the kinetic energy of the Compton electrons is converted to
the acceleration of the electrons which produced the observed radio
emission.  These must have $\gamma_e^\prime \sim 100$, making them many times
more energetic than the Compton electrons.
The interpenetration of Compton- and counter-currents is unstable to
electrostatic and electromagnetic plasma instabilities, which may explain
both the electron acceleration and the magnetic field (although previous
sections have also discussed Compton current, hydrodynamic dynamo and
pulsar radiation zone models of the field).  The absence of radio
polarization (\cite{K98}) suggests that the magnetic field is disordered,
which would be consistent with the dynamo, pulsar and plasma instability
models of the field, but might also be explained by differential Faraday
rotation (which may be estimated, using the previous parameters, as $\sim
10^4$ radians even at $\lambda = 3$ cm) within the Compton current model. 

Compton electron models predict the presence of the
nuclear gamma-rays.  At distances of many Mpc these gamma-rays are unlikely
to be detectable directly, but their effects on the visible light curve are
evident (and, in fact, led to the inference of $A = 56$ radioactive decay).  

In these models the rise of the radio flux in the first week is dependent on
the escape of the gamma-rays, which will only occur so soon for supernovae
(such as SN1998bw) with very fast debris.  It is therefore predicted that
strong early radio emission will be correlated with the debris expansion
speed.  It is also predicted that strong early radio emission requires the
presence of a surrounding medium, presumably the result of mass loss by the
progenitor, of density $> 10^4$ cm$^{-3}$, far in excess of typical
interstellar densities.  Such a medium may be independently confirmed or
disproved because of its effects on the SN visible light curve, pulse
dispersion of any newly born pulsar, or X-ray emission resulting from its
interaction with the debris.

As the debris collides with the radio-absorbing gas cloud a forward and
reverse shock are produced.  The forward shock has a post-shock temperature
$\sim 1$ MeV because of the high supernova debris speed, and emits
comparatively little radiation.  However, the reverse shock is cooler and
propagates in denser matter (the debris having a mass of several $M_\odot$,
compared to the mass $\ll M_\odot$ of the radio-absorbing envelope) and may
well produce the X-ray source S1 discovered (\cite{P99}) to be associated
with SN1998bw and to decay over $\sim 6$ months.  The decay is attributable
to the forward shock reaching the outer extent of the radio-absorbing cloud,
after which a rarefaction reflects and erodes the reverse shock.

Supernovae in which Compton electrons produce radio emission should have
several similar properties: They will all have approximately the same
Lorentz factors (1.6--2) for their expanding radio photospheres, because
this is determined by nuclear physics.  They will all show similar
double-peaked radio intensity curves, again because the nuclear physics is
the same.  Finally, their debris will show a nearly complete absence of a
low-Z envelope (as distinct from a surrounding low-Z medium)
because such an envelope would prevent the escape of the Compton electrons.

It is also worth noting that the energy of explosive C or O burning (about 1
MeV/nucleon) is insufficient to produce the debris velocity observed in
SN1998bw; gravitational energy from collapse to nearly neutron star density
(and probably neutrino transport) must be appealed to.

After the submission of the original version of this paper \cite{NP99}
reported that SN1987A was accompanied by two visible sources asymmetrically
located on opposite sides of the supernova.  Assuming these sources to
represent jets directed outward from the supernova at the same speed, they
found that their speed of ejection was $0.80 c$.  This is the speed of the
mean Compton electrons produced by the abundant 0.812 MeV gamma-rays of
$^{56}$Ni (\S 2), and is consistent with the radio source expansion speed
found by \cite{K98} for SN1998bw.  Although relativistic expansion is
observed at very different frequencies in these two objects, perhaps because
of observational limitations, it may be that both the spots near SN1987A and
the radio emission of SN1998bw are powered by similar Compton electrons.
\acknowledgements
I thank D. A. Frail and S. R. Kulkarni for useful discussions.

\end{document}